\documentclass[conference]{IEEEtran}
\IEEEoverridecommandlockouts
\usepackage{cite}

\usepackage[table]{xcolor}

\usepackage{amsmath,amssymb,amsfonts}
\usepackage{algorithmic}
\usepackage{graphicx}
\usepackage{textcomp}
\usepackage{xcolor}
\usepackage{multirow}
\def\BibTeX{{\rm B\kern-.05em{\sc i\kern-.025em b}\kern-.08em
    T\kern-.1667em\lower.7ex\hbox{E}\kern-.125emX}}

\begin{document}

\title{Continuous Speech Separation with Ad Hoc Microphone Arrays\\
}

\author{\IEEEauthorblockN{Dongmei Wang, Takuya Yoshioka, Zhuo Chen, Xiaofei Wang, Tianyan Zhou, Zhong Meng}
\IEEEauthorblockA{\textit{Microsoft, Redmond, WA, USA} \\
\{dowan, tayoshio, zhuc, xiaofewa, tizhou, zhme\}@microsoft.com}}

\maketitle

\begin{abstract}
Speech separation has been shown effective for multi-talker speech recognition. Under the ad hoc microphone array setup where the array consists of spatially distributed asynchronous microphones, additional challenges must be overcome  as the geometry and number of microphones are unknown beforehand. Prior studies show, with a spatial-temporal-interleaving structure, neural networks can efficiently utilize the multi-channel signals of the ad hoc array. In this paper, we further extend this approach to continuous speech separation. 
Several techniques are introduced to enable speech separation for real continuous recordings. 
First, 
we apply a transformer-based network for spatio-temporal modeling of the ad hoc array signals.    
In addition, two methods are proposed to mitigate a speech duplication problem during single talker segments, which seems more severe in the ad hoc array scenarios. One method is device distortion simulation for reducing the acoustic mismatch between simulated training data and real recordings.  The other is speaker counting to detect the single speaker segments and merge the output signal channels.
Experimental results for AdHoc-LibiCSS, a new dataset consisting of continuous recordings of concatenated LibriSpeech utterances obtained by multiple different devices, show the proposed separation method can significantly improve the ASR accuracy for overlapped speech with little performance degradation for single talker segments. 
\end{abstract}


\section{Introduction}

In multi-talker automatic speech recognition (ASR), speech separation plays a critical role for improving the recognition accuracy since conventional ASR systems cannot handle overlapped speech.  
While a microphone array with a known geometry has been widely used for far-field speech separation~\cite{ty_icassp_2018,Bahmaninezhad2019,MIMO-speech,zhongqiuwang_cmplx_css}, 
some attempts have recently been made to utilize ad hoc microphone arrays for speech separation and overlapped speech recognition~\cite{dongmei_2020, shota_2020, block_online, adhoc_extraction}.  
Compared with the fixed microphone array, the ad hoc microphone array comprising multiple independent recording devices, provides more flexibility and allows users to use their own mobile devices, such as cellphones or laptops, to virtually form the microphone array system. Moreover, the distributed devices can cover a wider space and thus provide more spatial diversity, which may be leveraged by the speech separation algorithms. 

There are two major challenges that arise from using the ad hoc arrays. One is the input permutation problem where the number and spatial arrangement of the microphones are unknown and unfixed. The other is that the individual microphone signals are asynchronous, 
which can be largely solved with cross-correlation-based approaches \cite{zliu_iwaenc_2008,Araki18-ICASSP,ty_interspeech_2019}. 
To handle the input permutation problem,  a spatial-temporal-interleaving (STI) neural network architecture was proposed~\cite{dongmei_2020}. This network models the spatial and temporal correlation by  stacking cross-channel self-attention layers and cross-frame BLSTM layers alternately. 
In \cite{shota_2020}, a guided source separation method was applied to the ad hoc array-based separation by using speaker diarization results, where a duplicate word reduction method was also proposed. 
In \cite{adhoc_extraction}, an ad hoc array-based target speech extraction was proposed by selecting 1-best or N-best channels for beamforming.   
Transform-average-concatenate~\cite{yiluo_icassp_2020} and 
a two stage-based method~\cite{nicolas_spatial_uncon_2020}
were proposed for spatially unconstrained microphone arrays, but they were only evaluated with simulated data. 

The previously proposed methods share a limitation that they require prior knowledge of utterance boundaries, which were often obtained from 
ground truth labels. However, in a realistic conversation scenario, the boundary information of overlapped speech is not easily obtainable. While \cite{shota_2020} used a speaker diarization system to acquire the utterance boundaries, it was based on offline processing whereas streaming processing is desired in many applications. 
In addition, in conversations, the speech overlap happens only occasionally. Therefore, the separation system must not only deal with the overlapped speech but also preserve the speech quality for single speaker regions so as not to degrade the ASR accuracy.

In this paper, we apply continuous speech separation (CSS) to the ad hoc microphone array setup. 
Previously, CSS was used for fixed microphone arrays~\cite{PrincetonASRU2019,chen2020continuous} and a single microphone setting~\cite{zhongqiuwang_cmplx_css,conformer_sep} to deal with real conversations. 
It outputs a fixed number (typically two) of audio channels, where each output channel contains at most one
active speaker at any time. 
When the input contains two overlapping utterances, CSS must separate them and emit the separated signals from different output channels. 
For segments with no speaker overlaps,
the incoming speech should be routed to one of the output
channels, while the other output channels produce zero or negligible
noise. 
For conversation transcription, a conventional recognition system can be simply applied to each output signal to enable multi-talker ASR. 

Three additional steps are proposed 
to address ad hoc array-based CSS challenges. 
A transformer-based architecture is adopted to model the spatial and temporal correlation of the ad hoc array signals. 
Moreover, two methods are introduced 
to mitigate the duplicate speech problem~\cite{shota_2020,zhongqiuwang_cmplx_css} in single speaker regions, which becomes severe especially when the array consists of different microphones.
One is based on data augmentation using device distortion simulation to mimic the acoustic variations of different devices and thereby reduce the mismatch between training data and real recordings.  
Also, speaker counting is applied to merge the CSS output channels into one if only one speaker is detected.    

To enable ad hoc array-based CSS evaluation, we collected a new dataset of long-form multi-talker audio with different consumer devices including cell phones and laptops, which we call AdHoc-LibriCSS. As with LibriCSS\cite{chen2020continuous}, LibriSpeech~\cite{panayotov2015librispeech} utterances were concatenated and played back in different conference rooms from multiple loudspeakers to create meeting-like audio files. 
Experimental results using this dataset are reported. 




\section{Continuous speech separation with ad hoc microphone arrays}
\label{sec:css}

\subsection{Continuous speech separation}
\label{ssec:css}

The CSS framework~\cite{ty_interspeech_2018,PrincetonASRU2019,chen2020continuous} attempts to cope with a long-form input signal including multiple partially overlapped or non-overlapped utterances in a streaming fashion. It is based on an observation that, most of the time, there are only one or two simultaneously active speakers in meeting conversations. 
CSS applies a sliding window to the input signal and performs separation within each window to produce a fixed number of separated signals (two in our experiments). The window size and the window shift we use are 4s and 2s, respectively. 
To make the output signal order consistent with that of the previous window position, 
the Euclidean distance is calculated between the separated signals of the current and previous windows over the overlapped frames between the two window positions for all possible output permutations. The output order with the lowest distance is then selected.
The separated signals are then concatenated with overlap-add technique.

\subsection{Transformer-based spatio-temporal modeling}
\label{ssec:transformer}

\begin{figure}[!t]
  \centering
  \centerline{\includegraphics[width=7.5cm]{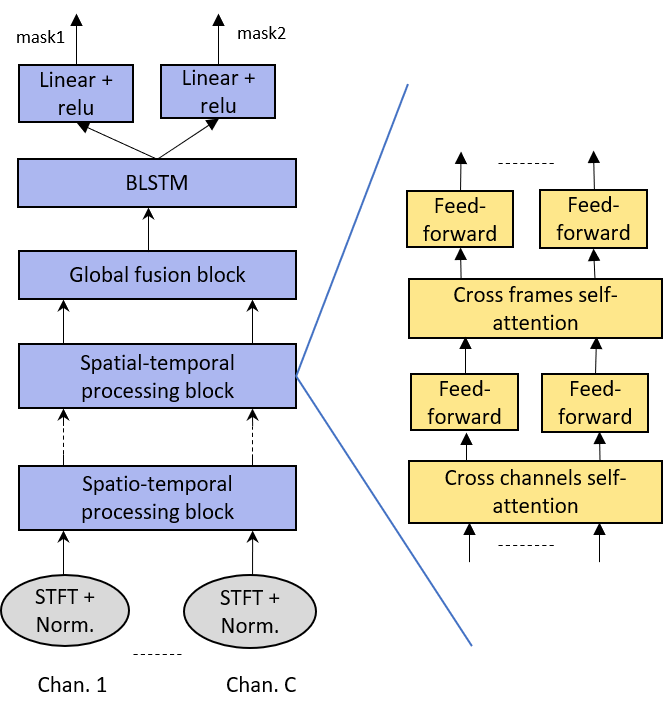}}
\caption{Overall separation model structure.}
\label{fig:overall_fig}
\vspace{-1em}
\end{figure}

Fig. \ref{fig:overall_fig} shows the overall architecture and the spatio-temporal processing block of our separation model.
The model consists of stacked spatio-temporal processing blocks, which adopts a transformer-based (or more precisely transformer encoder-based) architecture~\cite{transformer_2017}. 
The input to the separation model is a three-dimensional tensor comprising a multi-channel amplitude spectrogram, followed by global normalization \cite{dongmei_2020}.  
In the spatio-temporal processing block, 
a cross-channel self-attention layer exploits nonlinear spatial correlation between different channels and was shown effective in \cite{dongmei_2020}. A cross-frame self-attention layer allows the network to efficiently capture a long-range acoustic context~\cite{conformer_sep, self_attn_mc_2020, inter_SN_AAAI_2021}. 
After mean pooling-based global channel fusion, two BLSTM layers are further added to model the temporal correlation of the consolidated signals. Finally, two frequency domain masks are obtained with linear projection followed by ReLU activation.

\subsection{Channel selection}
\label{ssec:channel_selection}

In the ad hoc microphone array setting, the signal-to-noise ratio (SNR) may vary significantly across channels due to 
the differences in microphone characteristics as well as the large distances between different devices.
Therefore, the masks estimated for each speaker should be 
appplied to an appropriate channel. 
We perform  channel selection based on posterior SNR estimation~\cite{dongmei_2020,erdogan_interspeech_2016} for each CSS window. 
We directly apply the separation masks to the signals of the selected channels instead of enhansing the signals with mask-based beamforming~\cite{HEYMANN2017374,8461669}.
This is based on our informal observation that 
people often pick up their phones during meetings, making  beamforming challenging. 

\section{Addressing speech duplicating problem}

In real meetings, single speaker regions occupy most of the meeting time~\cite{cetin2006AnalysisOO}. Therefore, it is crucial for speech separation systems to preserve the audio quality for the single speaker regions while performing speech separation for the overlapped regions. 
Models trained with
permutation invariant training (PIT)~\cite{dyu_icassp_2017} tend to generate zero signals when there are fewer speakers than the model's output channels~\cite{ty_icassp_2018}. 
However, 
in the ad hoc microphone array settings, 
we observed that a resultant model still sometimes generated two output signals for a single speaker voice even when trained on both single- and multi-talker segments. 
This results in a high insertion error rate for ASR. 
This problem is more severe for the ad hoc microphone arrays as the same single speaker voice captured by 
different microphones can be acoustically very different. 
We describe
two methods for avoiding the duplicate speech problem: device distortion simulation and speaker counting. 

\subsection{Data augmentation with device distortion simulation}
\label{ssec:device_dis}

Device distortion simulation is a data augmentation scheme to reduce the mismatch between simulated training data and real multi-channel recordings obtained with different devices. 
The device distortion simulation consists of three steps: band-pass filtering, waveform amplitude clipping, and delay perturbation. 
Each step involves variable parameters, which are randomly chosen within a pre-set range for each microphone. The implementation details are described in Sec. \ref{ssec:training_data}.

\subsection{Output signal merger based on speaker counting}
\label{ssec:vad}

To further mitigate the speaker duplication issue, 
we apply speaker counting in each CSS processing window. 
When zero or one speaker is detected, 
the output signals of the separation model are merged 
into either one of the output channels by taking their sum. 
We then produce a zero signal from the other channel. 
The speaker counting is performed by using a randomly chosen one channel signal 
to avoid speaker counting errors caused by the data mismatch between multi-channel simulated training data and real recordings. 


A transformer-BLSTM model similar to the speech separatio model 
is trained for speaker counting. The model structure is the same as Fig. \ref{fig:overall_fig} except that 
the speaker counting model does not have cross-channel self-attention layers as it is based on a single channel input. 
The model input is an STFT of a randomly chosen single-channel signal. 
The model generates a frame-level speaker counting signal. 
We examine two output types for speaker counting. 
One model, which we call s1 in the experiment section, has 
a two-output linear layer followed by sigmoid nonlinearity for voice activity detection (VAD) for each speaker. 
One node gets activated when only one speaker is talking while two nodes become active when two people are speaking simultaneously.
This model is similar to the method proposed in \cite{zhongqiuwang_cmplx_css} and can be trained with PIT.  
Another model, which refers to as s2, has one linear output node for directly estimating the number of active speakers (0, 1 or 2 in our work).   
In both cases, we also add speech separation nodes and perform multi-task learning, which might
help better align the speaker counting learning with speech separation. 

For each CSS processing window, we determine whether there are multiple speakers in the currently processed window based on the model output and a predetermined threshold. 
For model s1, we decide that the current window contains multiple speakers if the two nodes get activated ($>0.5$) in three or more consecutive frames. 
For model s2, the criterion is whether 
the speaker counting node value is greater than 1.2 in three or more consecutive frames. 




\section{Experiment and results}
\label{sec:typestyle}

\subsection{Evaluation data}
\label{sec:dataset}
Following the development of LibriCSS \cite{chen2020continuous}, we designed and recorded a new dateset, namely AdHoc-LibiCSS, for evaluation of ad hoc array-based speech separation and multi-talker speech recognition algorithms under acoustically realistic conditions. The AdHoc-LibriCSS consists of recordings of concatenated LibriSpeech utterances played back from loudspeakers to simulate conversations. The recordings were made with multiple devices such as cell phones and laptops.

As with LibriCSS, the new dataset comprises multiple mini-sessions. Two different recording conditions are considered, which we refer to as 2-speaker and 5-speaker scenarios. The details of these two recording conditions are shown in Table \ref{tab:libricss_adhoc}.   
There are four subsets, dev-no-overlap, dev-overlap, test-no-overlap, and test-overlap, where the dev-$\ast$ and test-$\ast$ subsets use the LibriSpeech dev-clean and test-clean utterances, respectively. 
To enable fair comparison between the overlap and no-overlap conditions, the same speech content is used to create the overlap and no-overlap subsets. 

For each mini-session, we firstly sampled $N \in\{2, 5\}$ speakers from the LibriSpeech dev or test set \cite{panayotov2015librispeech} while ensuring that each utterance from every speaker was used only once in the recording. We then re-arranged and concatenated the utterances from each sampled speaker to form a simulated conversation, which was played by $N$ loudspeakers placed in a room. Each loudspeaker uniquely represented one talker. The loudspeakers and recording devices were randomly placed in the room. The setup remained the same within each mini-session.
The overlap ratio for test-overlap was in the range of $0$ to $30\%$, and that for dev-overlap was $10\%$ to $40\%$.
For each mini-session, 
all raw recordings from different devices were synchronized using cross-correlation before separation.


\begin{table}[t]
\centering
\caption{Recording setup details. 
}
\vspace{-.7em}
\label{tab:libricss_adhoc}
\begin{tabular}{lcc}
\hline
\hline
                      & 2-speaker  & 5-speaker \\ \hline
\#loudspeakers       & 2       & 5            \\room       & personal office       & meeting room            \\
duration per mini-session & 4 mins & 10 mins  \\
\#subsets / \#mini-sessions per subset & $4 / 20$ & $4 / 8$ \\
\#recording  devices & 5     & 5         \\ \hline
\end{tabular}
\end{table}


\subsection{Training data}
\label{ssec:training_data}
A training set consisting of 375 hours of artificially mixed speech was constructed for speech separation and speaker counting model training. 
We divided the training data into five categories based on the overlap style as proposed in \cite{ty_icassp_2018}: $40\%$ for single speaker segments, $9\%$ for inclusive overlap segments, $6\%$ for sequential overlap segments, $36\%$ for full overlap segments, and $9\%$ for partial overlap segments. 
Speaker and microhone locations as well as room dimensions were randomly determined to simulate the ad hoc array setting as described in \cite{dongmei_2020}, where room impulse responses were generated with the image method \cite{gpu_rir}.
Gaussian noise was added to each channel
at an SNR of $[-5, 15]$ dB.  
Device distortion simulation was then applied to the noisy overlapped signals. Each type of distortion was independently applied to each device. 
The band-pass filtering, waveform clipping, and delay perturbation were performed at probabilities of $40\%$, $5\%$, and $80\%$, respectively. The low and high cutoff frequencies of the band pass filter were uniformly sampled from $[50, 200]$ Hz and $[4000, 7000]$ Hz, respectively. The clipping ratio was uniformly sampled from $[0.55, 0.9]$. The delay for each device was uniformly sampled from $[-20, 20]$ ms. A validation set of $20$ hours was also generated in the same way.

\subsection{Training schemes}
\label{ssec:training}
For a separation model, the input waveform of each channel was transformed into an STFT representation with $257$ frequency bins every $16$ ms. Layer normalization was performed on the input magnitude spectrum vectors. Three spatio-temporal processing blocks were stacked. The self-attention for spatial modeling and temporal modeling both had $128$-dimensional embedding spaces and eight attention heads. The last two BLSTM layers contained $512$ cells for each direction.
We adopted PIT using an amplitude spectrum-based MSE loss.
The model was trained for $50$ epochs while saving the model parameters at the end of each epoch. 
The best model parameters were chosen based on the dev set WER. 

Our speaker counting models had three cross-frame self-attention layers, each followed by a feed-forward layer. 
Two BLSTM layers and a final linear layer are stacked on top.
The VAD-based s1 model had a sigmoid activation function to produce two VAD signals. 
For both models, we performed multi-task learning by using speech separation as an auxiliary task. 
It should be noted that, for s1 model training, PIT was independently applied to speech separation and VAD estimation. 
Both speaker counting models adopted an MSE loss for training.
The separation loss and the speaker counting loss were given an equal weight. 
At test time, the separation output was ignored. 
Model training was continued until a validation loss did not decrease for 10 continuous epochs. 


\begin{table}[t]
\centering
\caption{WERs of 2-speaker scenario (\%). Shaded and unshaded results are for the no-overlap and overlap subsets, respectively.}
\vspace{-.7em}
\label{tab:WER-2spk}
\begin{tabular}{ccccccc}
\hline \hline
\multirow{2}{*}{\begin{tabular}[c]{@{}c@{}}Overlap\\ ratio \%\end{tabular}} &
  \multirow{2}{*}{ori} &
  \multirow{2}{*}{sep} &
  \multirow{2}{*}{\begin{tabular}[c]{@{}c@{}}sep\\ +dis\end{tabular}} &
  \multicolumn{3}{c}{sep+dis+spk-cnt} \\ \cline{5-6} 
           &       &       &         & s1    & s2             \\ \hline
\multicolumn{7}{c}{dev-set}   \\ \hline
\rowcolor{gray!20}
0         & \textbf{12.01} & 18.13 & 16.02 & 13.23 & 12.46          \\
{[}10, 20) & 16.25 & 21.52 & 19.83         & 15.68 & \textbf{15.12}          \\
(20, 30) & 24.19 & 16.75 & 15.87      & 17.12 & \textbf{15.96} \\
(30, 40{]} & 32.65 & 20.98 & 19.85         & 20.14 & \textbf{19.64} \\ \hline
\multicolumn{7}{c}{test-set}   \\ \hline
\rowcolor{gray!20}
0         & 12.23 & 26.94 & 16.49 & 12.79 & \textbf{11.67}          \\
(0, 10)  & 12.93 & 17.88 & 13.48 & \textbf{12.08} & 12.25          \\
(10, 20) & 16.98 & 23.40 & 18.05          & 16.00 & \textbf{13.99} \\
(20, 30{]} & 25.08 & 29.86 & 18.85          & 16.24 & \textbf{15.17}
\\ \hline
\end{tabular}
\end{table}

\subsection{Evaluation scheme}
\label{ssec:asr}

For each mini-session, the CSS module using the trained separation model generated two output streams, each of which was then processed by a speech recognizer. Then, the recognition outputs were evaluated with asclite \cite{asclite_2006, sctk}, which can align multiple (two in this work) hypotheses against multiple reference transcriptions. 
We used an in-house hybrid ASR system~\cite{7953176} with 5-gram decoding trained on 33k hours of audio, including close-talking, distance-microhpone, and artificially corrupted speech. 



\subsection{Results and discussions}
\label{ssec:results}

Tables \ref{tab:WER-2spk} and \ref{tab:WER-5spk} show the WER results for various overlap ratios for the 2-speaker and 5-speaker scenarios, respectively.
For the dev-overlap and test-overlap subsets, the results are broken down by the mini-session overlap ratio. 
For each setting, we present the results of the following systems: (1) ASR applied to a randomly chosen channel without speech separation (ori); (2) ASR applied to the signals separated by the model trained without data augmentation (sep); (3) ASR applied to the signals separated by the model trained on device distortion simulated data (sep+dis); (4) systems performing speaker counting-based channel merger on top of (3) (sep+dis+spk-cnt). 


The results show that the separation model improved the WER for highly overlapped cases, but it resulted in significant degradation for less overlapped cases without the proposed duplication mitigation methods. 
This was mostly due to increased insertion errors.
Applying the device distortion simulation for the training data substantially improved the WERs in most cases.
However, the WER degradation for the no-overlap subsets was still significant for both the 2-speaker and 5-speaker cases. 
The channel merger processing using speaker counting mostly solved this problem, resulting in significant WER improvement for the highly overlapped data without compromising the ASR accuracy for the no-overlap subset. 
Among the two speaker counting schemes, the s2 system outperformed the s1 system in the 2-speaker scenario for almost all overlap conditions. 
In the 5-speaker case, both models performed equally well. 



\begin{table}[t]
\centering
\caption{WERs of 5-speaker scenario (\%). Shaded and unshaded results are for the no-overlap and overlap subsets, respectively.}
\vspace{-.7em}
\label{tab:WER-5spk}
\begin{tabular}{ccccccc}
\hline \hline
\multirow{2}{*}{\begin{tabular}[c]{@{}c@{}}Overlap\\ ratio \%\end{tabular}} &
  \multirow{2}{*}{ori} &
  \multirow{2}{*}{sep} &
  \multirow{2}{*}{\begin{tabular}[c]{@{}c@{}}sep\\ +dis\end{tabular}} &
  \multicolumn{3}{c}{sep+dis+spk-cnt} \\ \cline{5-6} 
           &       &       &              & s1             & s2             \\ \hline
\multicolumn{7}{c}{dev-set}   \\ \hline
\rowcolor{gray!20}
0          & 12.85 & 17.19 & 15.50          & 13.05          & \textbf{12.15} \\ 
{[}10, 20) & 16.80 & 17.03 & 15.47  & \textbf{13.69}          & 13.90          \\
(20, 30) & 26.38 & 16.76 & 18.58  & \textbf{16.50}          & 16.73          \\
(30, 40{]} & 28.28 & 19.09 & 19.51          & 18.64          & \textbf{17.68} \\ \hline
\multicolumn{7}{c}{test-set}   \\ \hline
\rowcolor{gray!20}
0          & 15.50  & 20.44 & 16.57 & \textbf{13.62}          & 13.70          \\
(0, 10)  & 15.27 & 15.60 & 12.79    & 11.51          & \textbf{11.38}          \\
(10, 20) & 21.07 & 16.76 & 15.65    & \textbf{14.82} & 15.43          \\
(20, 30{]} & 29.42 & 22.27 & 20.05  & 17.92          & \textbf{17.34}          \\ \hline
\end{tabular}
\end{table}



\section{Conclusions}
\label{sec:conclusions}

We described a CSS system for ad hoc microphone arrays.
A transformer-based architecture was applied for separation. 
To mitigate the speech duplicating problem for non-overlapped segments, we proposed data augmentation based on device distortion simulation 
to reduce the mismatch between training data and the real recordings obtained with spatially distributed devices. 
The use of speaker counting was also introduced to further mitigate the issue.
Multi-talker ASR experiments were performed by using newly recorded AdHoc-LibriCSS,
showing that the proposed system significantly improved the ASR accuracy for recordings including various degrees of overlaps while retaining the WER for non-overlapped speech. 

\newpage


\bibliographystyle{IEEEbib}
\bibliography{strings,refs}

\end{document}